\documentclass[aps,preprint,nofootinbib,showpacs]{revtex4}
\usepackage[dvips]{graphicx,color}
\usepackage{amsmath}
\newcommand{\be}{\begin{equation}}
\newcommand{\ee}{\end{equation}}
\newcommand{\ba}{\begin{eqnarray}}
\newcommand{\beq}{\begin{equation}}
\newcommand{\eeq}{\end{equation}}
\newcommand{\ea}{\end{eqnarray}}

\newcommand{\MeV}{\text{MeV}}
\newcommand{\GeV}{\text{GeV}}
\newcommand{\TeV}{\text{TeV}}
\newcommand{\fb}{\text{fb}}

\newcommand{\mll}{{m_{\ell\ell^\prime}^{}}}
\newcommand{\hll}{{h_{\ell\ell^\prime}^{}}}
\newcommand{\qqHppHmm}{{q\overline{q}\to \gamma^*,Z^* \to H^{++}H^{--}}}

\newcommand{\qqHpmpmHmp}{{q^\prime\overline{q}\to W^* \to H^{\pm\pm}H^\mp}}

\newcommand{\Hpmlpmnu}{{H^\pm \to \ell^\pm \nu_{\ell^\prime}^{}}}

\def\beqa{\begin{eqnarray}}
\def\eeqa{\end{eqnarray}}

\def\bea{\begin{eqnarray}}
\def\eea{\end{eqnarray}}

\def\err#1#2{\lower2pt\hbox{ $\stackrel{\scriptstyle +#1}{\scriptstyle -#2}$}}
\def\ga{\mathrel{\raise.3ex\hbox{$>$\kern-.75em\lower1ex\hbox{$\sim$}}}}
\def\la{\mathrel{\raise.3ex\hbox{$<$\kern-.75em\lower1ex\hbox{$\sim$}}}}
\def\bmaT{\left(\begin{array}{ccc}}
\def\emaT{\end{array}\right)}
\def\bma{\left( \begin{array} }
\def\ema{\end{array} \right)}
\def\gsim{~{\rlap{\lower 3.5pt\hbox{$\mathchar\sim$}}\raise 1pt\hbox{$>$}}\,}
\def\lsim{~{\rlap{\lower 3.5pt\hbox{$\mathchar\sim$}}\raise 1pt\hbox{$<$}}\,}

\begin{document}

\preprint{
\vbox{%
\hbox{SHEP-12-16}
}}
\title{\boldmath Enhancement of $H\to \gamma\gamma$ from doubly charged scalars\\
in the Higgs Triplet Model \unboldmath} 
\author{A.G. Akeroyd}
\email{a.g.akeroyd@soton.ac.uk}
\author{S. Moretti}
\email{s.moretti@soton.ac.uk}
\affiliation{School of Physics and Astronomy, University of Southampton, \\
Highfield, Southampton SO17 1BJ, United Kingdom,}
\affiliation{Particle Physics Department, Rutherford Appleton Laboratory, Chilton, Didcot, 
Oxon OX11 0QX, United Kingdom}

\date{\today}
\begin{abstract}

The Higgs boson of the Standard Model is being searched for at the LHC 
in the channel $H\to \gamma\gamma$. In the Higgs Triplet Model (HTM) there are contributions 
to this decay from loops of doubly charged scalars ($H^{\pm\pm}$) and
singly charged scalars $(H^\pm)$ that are not present in the Standard Model. These additional 
contributions are mediated by the trilinear couplings $H_1 H^{++}H^{--}$ and $H_1 H^{+}H^{-}$, 
where $H_1$ is the lightest CP-even Higgs boson.
We point out the possibility of constructive interference of the $H^{\pm\pm}$ contribution
with that of the $W^\pm$ contribution, which enables a 
substantial enhancement of the branching ratio of $H_1\to \gamma\gamma$  in the HTM.
The magnitude of the contribution of $H^{\pm\pm}$ is essentially determined by
the mass of $H^{\pm\pm}$ ($m_{H^{\pm\pm}}$) and a quartic scalar coupling ($\lambda_1$), with
constructive interference arising for $\lambda_1 < 0$. 
Consequently, the ongoing searches for $H\to \gamma\gamma$ restrict the parameter space of
$[m_{H^{\pm\pm}},\lambda_1]$ more stringently for $\lambda_1 < 0$ than for
the recently-studied case of destructive interference with $\lambda_1 > 0$.
Moreover, if the excess of $\gamma\gamma$ events around a mass of 125 GeV in the LHC searches for  $H\to \gamma\gamma$
is substantiated with larger data samples, and the branching ratio is measured to be somewhat larger than that
for the SM Higgs boson, then such an enhancement could be readily accommodated by $H_1$ in the HTM with $\lambda_1 <0$.

\end{abstract}
\maketitle


\section{Introduction} 
\noindent
There is much ongoing experimental effort by the ATLAS and CMS collaborations at the CERN 
Large Hadron Collider (LHC)
to search for the neutral Higgs boson ($h^0$) of the Standard Model (SM) \cite{ATLAS:2012ae,Chatrchyan:2012tx,latest_LHC}.
This model of spontaneous symmetry breaking  will be tested at the 
LHC over all of the theoretically preferred mass range,
in an experimental programme which is expected to be completed by the end of the $\sqrt s=8$ TeV run of the LHC.
After analysing all of the data taken at $\sqrt s=7$ TeV \cite{latest_LHC} there are only two regions for the mass of
$h^0$ which have not been excluded at 95\% c.l:
 i) a region of light mass, roughly corresponding to $122 \,{\rm GeV} < m_{h^0} \,< 128 \,{\rm GeV}$, and ii) a region
of heavy mass, $m_{h^0}> 600$ GeV. 
At present there is much speculation about an excess of events suggesting a mass of around 125 GeV, 
especially in the channel $H\to \gamma\gamma$ 
\cite{Chatrchyan:2012tw,ATLAS:2012ad} where
the excess is a bit larger than that expected for a SM Higgs boson of the same mass. More data is needed to
clarify if this small excess is a fluctuation of the background, or if it is genuine production of a Higgs boson.
Several recent studies 
\cite{Carmi:2012yp, Gabrielli:2012yz} fit the current data in all the Higgs search channels
to the case of a neutral Higgs boson with arbitrary couplings. A SM-like Higgs boson gives a good fit to the
data, although a slight preference for non-SM like Higgs couplings is emphasised in \cite{Gabrielli:2012yz,Giardino:2012ww}.

The Higgs sector of the SM, which consists of one fundamental scalar with a 
vacuum expectation value (vev), might not be nature's choice. There could be
additional scalar fields which also contribute to the masses of the fermions and weak bosons, with a 
more complicated scalar potential which depends on several arbitrary parameters.  
Importantly, even in the event of no signal for a SM-like Higgs boson at the LHC, the search for
scalar particles should continue in earnest due to the fact that a non-minimal Higgs sector
can give rise to different experimental signatures, some of which are challenging to detect.
Consequently, it will take much longer for the LHC to probe all of the parameter space of such models.
Various interpretations of the excess of events at around 125 GeV have been proposed in the context of models with two Higgs doublets
in \cite{Ferreira:2011aa}, with a review of such models in
\cite{Branco:2011iw}. Early studies of non-minimal Higgs sectors which can give rise to an enhanced BR$(H\to \gamma\gamma)$
can be found in \cite{Arhrib:2004ak} 
(in which the enhancement is caused by increasing the partial width of $H\to \gamma\gamma$) and in \cite{Barger:1992ty}
(in which the dominate tree-level decays modes such as $H\to b\overline b$ are suppressed.)

The Higgs Triplet Model (HTM) \cite{Konetschny:1977bn,Mohapatra:1979ia,Magg:1980ut,Schechter:1980gr,Cheng:1980qt}
is a model of neutrino mass generation with a non-minimal Higgs sector.
The model predicts several scalar particles, 
including a doubly charged Higgs boson ($H^{\pm\pm}$) and a singly charged Higgs boson ($H^{\pm}$), for which
direct searches are being  carried out at the LHC \cite{CMS-search,Aad:2012cg}. In a large
part of the parameter space of the HTM the lightest CP-even scalar ($H_1$) has essentially the same couplings 
to the fermions and vector bosons as the Higgs boson of the SM \cite{Dey:2008jm,Akeroyd:2010je,Arhrib:2011uy}.
Therefore the ongoing searches for the SM Higgs boson also apply to $H_1$ of the HTM with very little modification.
An exception is the loop-induced decay $H_1\to \gamma\gamma$ which receives
contributions from virtual $H^{\pm\pm}$ and $H^{\pm}$, and can have a branching ratio which is very
different to that of the SM Higgs boson. As shown recently in \cite{Arhrib:2011vc},
the ongoing limits on BR($H_1\to \gamma\gamma$) constrain the parameter space of
$[m_{H^{\pm\pm}},\lambda_1]$, where $\lambda_1$ is
a quartic coupling in the scalar potential (see also \cite{Kanemura:2012rs} for a related study). The case of $\lambda_1 > 0$ was studied in
\cite{Arhrib:2011vc}, which leads to destructive interference between the 
combined SM contribution (from $W$ and fermion loops) and the contribution from $H^{\pm\pm}$.
In this work we consider the case of $\lambda_1 < 0$, which leads to constructive interference and was not
considered in \cite{Arhrib:2011vc,Kanemura:2012rs}.
The scenario of $\lambda_1 < 0$ is more constrained by the ongoing searches for $H_1\to \gamma\gamma$ than the case of 
$\lambda_1 > 0$. Moreover, the scenario of $\lambda_1 < 0$ can provide enhancements of
 $H_1\to \gamma\gamma$ with smaller $|\lambda_1|$ than the case of $\lambda_1 > 0$.

Our work is organised as follows. In section~II we briefly describe
the theoretical structure of the HTM. In section III we present the formulae for
the contribution of $H^{\pm\pm}$ and $H^{\pm}$ to $H_1\to \gamma\gamma$, and we 
summarise the ongoing searches for these particles at the LHC. Section IV contains our numerical results,
with conclusions in section V.

\section{The Higgs Triplet Model}

 In the HTM~\cite{Konetschny:1977bn,Schechter:1980gr,Cheng:1980qt}
a $Y=2$ complex $SU(2)_L$ isospin triplet of scalar fields,
${\bf T}=( T_1, T_2, T_3 )$, is added to the SM Lagrangian. 
 Such a model can provide Majorana masses for the observed neutrinos 
without the introduction of $SU(2)_L$ singlet neutrinos
via the gauge invariant Yukawa interaction:
\begin{equation}
{\cal L}=\hll L_\ell^TCi\tau_2\Delta L_{\ell^\prime}+\text{h.c.}
\label{trip_yuk}
\end{equation}
 Here $\hll (\ell,\ell^\prime=e,\mu,\tau)$ is a complex
and symmetric coupling,
$C$ is the Dirac charge conjugation operator,
$\tau_i (i=1\text{-}3)$ are the Pauli matrices,
$L_\ell=(\nu_{\ell L}, \ell_L)^T$ is a left-handed lepton doublet,
and $\Delta$  is a $2\times 2$ representation
of the $Y=2$ complex triplet fields:
\begin{equation}
\Delta
= {\bf T\cdot\tau}
= T_1 \tau_1 + T_2 \tau_2 + T_3 \tau_3
=\bma{cc}
\Delta^+/\sqrt{2}  & \Delta^{++} \\
\Delta^0       & -\Delta^+/\sqrt{2}
\ema ,
\end{equation}
where
$T_1 = (\Delta^{++} + \Delta^0)/2$,
$T_2 = i(\Delta^{++} - \Delta^0)/2$,
and $T_3 = \Delta^+ / \sqrt{2}$.
A non-zero triplet vev $\langle\Delta^0\rangle$ 
gives rise to the following mass matrix for neutrinos:
\begin{equation}
\mll = 2\hll \langle\Delta^0\rangle = \sqrt{2}\hll v_{\Delta} .
\label{nu_mass}
\end{equation}
The necessary non-zero $v_{\Delta}$ arises from the minimisation of
the most general $SU(2)_L\otimes U(1)_Y$ invariant Higgs potential~\cite{Cheng:1980qt,Gelmini:1980re},
which is written
as follows~\cite{Ma:2000wp, Chun:2003ej}
(with $H=(\phi^+,\phi^0)^T$):

\begin{eqnarray}
V(H,\Delta) & = & 
- m_H^2 \ H^\dagger H \ + \ \frac{\lambda}{4} (H^\dagger H)^2 \ 
+ \ M_{\Delta}^2 \ {\rm Tr} \Delta^\dagger \Delta\ 
+ \ \left( \mu \ H^T \ i \tau_2 \ \Delta^\dagger H \ + \ {\rm h.c.}\right) \ 
\nonumber \\
&& + \ \lambda_1 \ (H^\dagger H) {\rm Tr} \Delta^\dagger \Delta \ 
+ \ \lambda_2 \ \left( {\rm Tr} \Delta^\dagger \Delta \right)^2 \ 
+ \ \lambda_3 \ {\rm Tr} \left( \Delta^\dagger \Delta \right)^2 \ 
+ \ \lambda_4 \ H^\dagger \Delta \Delta^\dagger H. 
\label{Potential}
\end{eqnarray}

Here $m^2<0$
in order to ensure non-zero $\langle\phi^0\rangle=v/\sqrt 2$
which spontaneously breaks
$SU(2)_L \otimes U(1)_Y$ to $U(1)_Q$
while $M^2_\Delta > 0$.
The scalar potential in eq.~(\ref{Potential})
together with the triplet Yukawa interaction of eq.~(\ref{trip_yuk})
lead to a phenomenologically viable model
of neutrino mass generation.
 For small $v_\Delta/v$,
the expression for $v_\Delta$
resulting from the minimisation of $V$ is:
\begin{equation}
v_\Delta
\simeq
 \frac{ \mu v^2 }
      { \sqrt{2} ( M^2_\Delta + v^2 (\lambda_1+\lambda_4)/2 ) }\,. 
\label{tripletvev}
\end{equation}

 For $M_\Delta \gg v$
one has $v_\Delta \simeq \mu v^2/(2M^2_\Delta)$,
which would naturally lead to a small $v_\Delta$
even for $\mu$ of the order of the electroweak scale
(and is sometimes called the ``Type II seesaw mechanism'').
 However, in this scenario the triplet scalars would be too heavy to be observed at the LHC\@. 
In recent years there has been much interest
in  the case of light triplet scalars ($M_\Delta\approx v$) 
within the discovery reach of the LHC,
for which eq.~(\ref{tripletvev}) leads to $v_\Delta\approx \mu$, and 
this is the scenario we will focus on.
The case of $v_\Delta < 0.1\,\MeV$
is assumed in the ongoing searches at the LHC, 
for which the BRs of the triplet scalars
to leptonic final states
(e.g.\ $H^{\pm\pm}\to \ell^\pm\ell^\pm$) would be $\sim 100\%$. 
Since $v_\Delta\approx \mu$ for 
light triplet scalars then $\mu$ must also be small 
(compared to the electroweak scale) for the scenario of 
$v_\Delta < 0.1\,\MeV$. Moreover,
if one requires
that the triplet Yukawa couplings $h_{ij}$ are
greater in magnitude than the smallest Yukawa coupling in the SM
(i.e.\ the electron Yukawa coupling, $y_e\sim 10^{-6}$)
then from eq.~(\ref{nu_mass})
one has $v_\Delta < 0.1\,\MeV$,
and thus the decays of  the triplet scalars to leptonic final states
have BRs which sum to $\sim 100\%$.
In extensions of the HTM
the term $\mu(H^Ti\tau_2\Delta^\dagger H$), which is the only source of 
explicit lepton number violation, may arise in various ways:
 i) it can be generated at tree level
via the vev of a Higgs singlet field~\cite{Schechter:1981cv}; 
 ii) it can arise at higher orders
in perturbation theory~\cite{Chun:2003ej};
 iii) it can originate
in the context of extra dimensions~\cite{Ma:2000wp}, and iv) in the context of
other extensions of the HTM \cite{Majee:2010ar,Kanemura:2012rj}.
 
The branching ratios~(BRs) for 
$H^{\pm\pm}\to \ell^\pm{\ell^\prime}^\pm$ depend on $\hll$
and are predicted in the HTM in terms of the parameters
of the neutrino mass matrix~\cite{Akeroyd:2005gt, Ma:2000wp, Chun:2003ej}.
 Detailed quantitative studies of
BR($H^{\pm\pm}\to \ell^\pm{\ell^\prime}^\pm$) in the HTM
have been performed in
Refs.~\cite{Garayoa:2007fw,Akeroyd:2007zv,Kadastik:2007yd,
Perez:2008ha,delAguila:2008cj,Akeroyd:2009hb} 
with particular emphasis  
given to their sensitivity to the Majorana phases and 
the absolute neutrino mass i.e.\ parameters which cannot be 
probed in neutrino oscillation experiments. 
 A study on the relation
between BR($H^{\pm\pm}\to \ell^\pm{\ell^\prime}^\pm$)
and the neutrinoless double beta decay
was performed in Ref.~\cite{Petcov:2009zr}.

An upper limit on $v_\Delta$ can be obtained
from considering its effect on the parameter
$\rho (=M^2_W/M_Z^2\cos^2\theta_W)$. 
 In the SM $\rho=1$ at tree-level,
while in the HTM one has (where $x=v_\Delta/v$):
\begin{equation}
\rho
\equiv
 1 + \delta\rho = {1+2x^2 \over 1+4x^2} .
\label{deltarho}
\end{equation}
 The measurement $\rho\approx 1$ leads to the bound
$v_\Delta/v\lsim 0.03$, or  $v_\Delta\lsim 8\,\GeV$. Therefore the vev of the doublet
field $v$ is essentially equal to the vev of the Higgs boson of the SM 
(i.e. $v\approx 246$ GeV). At the 1-loop level
$v_\Delta$ must be renormalised \cite{Blank:1997qa}, and the first explicit analysis
in the context of the HTM has recently been performed
in \cite{Kanemura:2012rs}, with bounds on  $v_\Delta$ similar to those of the 
tree-level analysis.

 The scalar eigenstates in the HTM are as follows:
 i) the charged scalars $H^{\pm\pm}$ and $H^\pm$;
 ii) the CP-even neutral scalars $H_1$ and $H_2$;
 iii) a CP-odd neutral scalar $A^0$.
 The doubly charged $H^{\pm\pm}$ is entirely composed
of the triplet scalar field $\Delta^{\pm\pm}$, 
while the remaining eigenstates are in general mixtures
of the doublet and triplet fields.
However,
such mixing is proportional to the triplet vev,
and hence small {\it even if} $v_\Delta$
assumes its largest value of a few GeV. The expressions for the
mixing angle in the CP-even sector $\alpha$ 
and charged Higgs sector $\beta'$ are:
\begin{equation}
\sin\alpha\sim  2v_\Delta/v\,,\,\,\   \sin\beta'=\sqrt 2 v_\Delta/v \, .   
\label{alpha_beta}
\end{equation}
We note that a large mixing angle (including maximal mixing, $\alpha=\pi/4)$ is possible in the CP-even sector
provided that $m_{H_1}\sim m_{H_2}$~\cite{Dey:2008jm,Akeroyd:2010je,Arhrib:2011uy}, but we will
not consider this parameter space in this work. The above approximation for $\sin\alpha$ is
valid as long as $m_{H_2}$ is larger than $m_{H_1}$ by 5 GeV or so. Therefore
$H^\pm,H_2,A^0$ are predominantly composed of the triplet fields,
while $H_1$ is predominantly composed of the doublet field
and plays the role of the SM Higgs boson.
Neglecting the small off-diagonal elements in the CP-even mass matrix, 
the approximate expressions for the squared masses of $H_1$ and $H_2$ are
as follows:
\begin{eqnarray}
m^2_{H_1}= 
\frac{\lambda}{2} v^2 \, , \\ 
\label{massH2}
m^2_{H_2}= M^2_\Delta + (\frac{\lambda_1}{2} 
 +\frac{\lambda_4}{2}) v^2 + 3(\lambda_2 +\lambda_3) v^2_\Delta \, .
\label{massH1}
\end{eqnarray}
The squared mass of the (dominantly triplet) CP-odd $A^0$ is given by:
\begin{equation}
m^2_{A^0} = M^2_\Delta + (\frac{\lambda_1}{2} 
 +\frac{\lambda_4}{2}) v^2 + (\lambda_2 +\lambda_3) v^2_\Delta \, .
\label{CP-odd-mass}
\end{equation}
The squared mass of the (dominantly triplet) $H^\pm$ is given by:
\begin{equation}
m^2_{H^\pm} = M^2_\Delta + (\frac{\lambda_1}{2} 
 +\frac{\lambda_4}{4}) v^2 + (\lambda_2  + \sqrt 2
 \lambda_3) v^2_\Delta \, .
 \qquad
\label{charged-mass}
\end{equation}
Finally, the squared mass of the (purely triplet) doubly-charged 
scalar ($H^{\pm\pm} = \delta^{\pm\pm}$) is given by:
\begin{equation}
m^2_{H^{\pm\pm}} =M^2_\Delta + \frac{\lambda_1}{2} v^2
+\lambda_2 v_\Delta^2 \, .
\label{doub-charged-mass}
\end{equation}
One can see that the squared mass of the (dominantly doublet) $H_1$
 is simply given by $\lambda v^2/2$, as in the
SM. In the expressions for the masses
of $m^2_{A^0}$, $m^2_{H_2}$, $m^2_{H^\pm}$ and $m^2_{H^{\pm\pm}}$ 
there is a common term $M^2_\Delta + \frac{\lambda_1}{2} v^2$.
It is evident that the mass scales for $H_1$ and
the dominantly triplet scalars ($A^0,H_2, H^\pm, H^{\pm\pm}$) are unrelated, the former being set 
by $\lambda v^2/2$ and the latter by $M^2_\Delta + \frac{\lambda_1}{2} v^2$.
Neglecting the terms which are proportional to the small parameter $v_\Delta$, one can see that 
there are only two possible mass hierarchies for the triplet scalars, with the magnitude of the
mass splitting being controlled by $\lambda_4$ (and $m_{A^0}=m_{H_1}$ when $v_\Delta$ is neglected):
\begin{eqnarray}
m_{A^0},m_{H_2} < m_{H^\pm} < m_{H^{\pm\pm}} \;\;{\rm for}\;\; \lambda_4<0 \, ,
\label{mass-split1} \\
m_{H^{\pm\pm}}  < m_{H^\pm} <  m_{A^0},m_{H_2} \;\; {\rm for} \;\; \lambda_4>0 \, .
\label{mass-split2}
\end{eqnarray}
In our numerical analysis we choose the following seven parameters
from the scalar potential to be the input parameters:
\begin{equation}
\lambda, \, \lambda_1, \, \lambda_2, \, \lambda_3,\, \lambda_4, \, v_\Delta \,, m_{H^{\pm\pm}} \, .
\end{equation}
Therefore $M^2_\Delta$ and $\mu$ are determined from the above parameters.

\section{The decay $H_1\to \gamma\gamma$ in the HTM}
\noindent
Many previous works have studied the impact of singly charged scalars on the
decay $H \rightarrow\gamma\gamma$ e.g. in the context of the minimal supersymmetric SM (MSSM) \cite{Djouadi:1996pb},
a Two-Higgs Doublet Model \cite{Arhrib:2004ak,Ginzburg:2001wj} and the Next-to-MSSM \cite{Ellwanger:2011aa}. 
The contribution of doubly charged scalars to this decay has received comparatively very little
attention.\footnote{See Ref.~\cite{Alves:2011kc} for a study of the effect of doubly charged vector bosons
on $H \rightarrow\gamma\gamma$ in the context of a 3-3-1 model.} A study in the context of
a Little Higgs Model was performed in \cite{Han:2003gf}, but the magnitude of the contribution
from $H^{\pm\pm}$ was shown to be
much smaller than that of the singly charged scalar in the same model, due to the theoretical structure of the
scalar potential.
Recently the impact of the contribution from $H^{\pm\pm}$ was studied in the HTM in \cite{Arhrib:2011vc,Kanemura:2012rs},
and was shown to give a sizeable contribution to $H \rightarrow\gamma\gamma$. In \cite{Arhrib:2011vc} both
enhancements and suppressions were discussed, while in \cite{Kanemura:2012rs} only suppressions were studied.
We will closely
follow the analysis of \cite{Arhrib:2011vc}. Note that
the loop-induced decay $H_1\to \gamma Z$ would also receive contributions from $H^{\pm\pm}$ and
 $H^{\pm}$, but we will not consider this decay in this work because it is not expected to
be observable in the early stages of operation of the LHC.
We will consider the parameter space where $H_1$ is essentially composed of the
doublet field, while $H_2$ is essentially composed of the neutral triplet field, which
is realised in a large region of the parameter space of the HTM. In this case
both $H_2$ and $A^0$ will be difficult to observe in the 8 TeV run of the LHC due to 
their small couplings to quarks and vector bosons, and so they can only be produced in reasonable amounts by mechanisms
such as $pp\to Z^*\to H^0A^0$ (as discussed in \cite{Akeroyd:2012nd}).
Consequently, the mass limits from the searches for the SM Higgs in the interval $110 \,{\rm GeV} < M_H < 600$ GeV do not apply
to $H_2$ and $A^0$ since their production cross sections are much smaller than that for $gg\to H$.

\subsection{Decay rate for $H_1\to \gamma\gamma$}
The explicit form of the decay width for $H_1 \rightarrow\gamma\gamma$ is as follows \cite{Shifman:1979eb}:
\begin{eqnarray}
\label{eq:DTHM-h2gaga}
\Gamma(H_1 \rightarrow\gamma\gamma)
& = & \frac{G_F\alpha^2 M_{H_1}^3}
{128\sqrt{2}\pi^3} \bigg| \sum_f N_c Q_f^2 g_{H_1 ff} 
A_{1/2}^{H_1}
(\tau_f) + g_{H_1 WW} A_1^{H_1} (\tau_W) \nonumber \\
&& + \tilde{g}_{H_1 H^\pm\,H^\mp}
A_0^{H_1}(\tau_{H^{\pm}})+
 4 \tilde{g}_{H_1 H^{\pm\pm}H^{\mp\mp}}
A_0^{H_1}(\tau_{H^{\pm\pm}}) \bigg|^2 \, .
\label{partial_width_htm}
\end{eqnarray}
Here $\alpha$ is the fine-structure constant, $N_c(=3)$ is the number of quark colours, $Q_f$ is the
electric charge of the fermion in the loop, and $\tau_{i}=m^2_{H_1}/4m^2_{i}$ $(i=f,W,H^{\pm},H^{\pm\pm})$. The loop functions $A_1$ (for the $W$ boson) and $A_{1/2}$
(for the fermions, $f$) are well known. For the contribution from the fermion loops we will only keep the term with the top quark,
which is dominant. The loop function for $H^{\pm\pm}$ and $H^{\pm}$ is given by:
\begin{eqnarray}
A_{0}^{H_1}(\tau) &=& -[\tau -f(\tau)]\, \tau^{-2} \, ,
\label{eq:Ascalar}
\end{eqnarray}
and the function $f(\tau)$ is given by
\begin{eqnarray}
f(\tau)=\left\{
\begin{array}{ll}  \displaystyle
\arcsin^2\sqrt{\tau} & \tau\leq 1 \\
\displaystyle -\frac{1}{4}\left[ \log\frac{1+\sqrt{1-\tau^{-1}}}
{1-\sqrt{1-\tau^{-1}}}-i\pi \right]^2 \hspace{0.5cm} & \tau>1 \, .
\end{array} \right. 
\label{eq:ftau} 
\end{eqnarray}
Note that the contribution from the loop with $H^{\pm\pm}$ in eq.(\ref{partial_width_htm})
is enhanced relative to the contribution from $H^{\pm}$ by a factor of four at the amplitude level.
The couplings of $H_1$ to the vector bosons and fermions relative to the values in the SM are as follows:
\begin{eqnarray}
g_{H_1 t\overline t}=\cos\alpha/\cos\beta' \, ,
\label{h1tt}\\
g_{H_1 b\overline b}=\cos\alpha/\cos\beta' \, , 
\label{h1bb} \\
g_{H_1WW}= \cos\alpha+2\sin\alpha v_\Delta/v  \, ,  
\label{h1WW} \\
g_{H_1ZZ}= \cos\alpha+4\sin\alpha v_\Delta/v \, .
\label{h1ZZ}
\end{eqnarray}
From Eq~.(\ref{alpha_beta}) one has $\cos\alpha \sim \sqrt{(1-4v_\Delta^2/v^2)} \sim 1$, 
and  $\cos\beta'=\sqrt{(1-2v_\Delta^2/v^2}) \sim 1$, and so it is clear that the above couplings
of $H_1$ are essentially identical to those of the SM Higgs boson because $v_\Delta<< v$.

The scalar trilinear couplings are parametrised as follows:
\begin{eqnarray}
\tilde{g}_{H_1 H^{++}H^{--}}  & = & -  \frac{m_W}{ g m_{H^{\pm \pm}}^2} g_{H_1 H^{++}H^{--}} \, ,
 \label{eq:redgcalHHpp}\\
\tilde{g}_{H_1 H^+H^-} & = & -  \frac{m_W}{g m_{H^{\pm}}^2} g_{H_1 H^+H^-} \, ,
\label{eq:redgcalHHp}
\end{eqnarray}
with the following explicit expressions in terms of the parameters of the scalar potential
(where $s_\alpha=\sin\alpha$ etc) \cite{Arhrib:2011uy}:
\begin{equation}
g_{H_1H^{++}H^{--}} =-\{2\lambda_2v_\Delta s_\alpha+\lambda_1v c_\alpha\} \, ,
 \label{eq:ghHpp}
\end{equation}
\begin{eqnarray}
g_{H_1H^+H^-}=-\frac{1}{2}
\bigg\{\{4v_\Delta(\lambda_2 + \lambda_3) c_{\beta'}^2+2v_\Delta\lambda_1s_{\beta'}^2-
\sqrt{2}\lambda_4v c_{\beta'}s_{\beta'}\}s_\alpha  \nonumber \\ 
+\{\lambda\,v s_{\beta'}^2+{(2\lambda_{1}+\lambda_{4}) }v c_{\beta'}^2+
(4\mu-\sqrt{2}\lambda_4v_\Delta)c_{\beta'}s_{\beta'}\}c_\alpha\bigg\} \, .
\label{eq:ghHp}
\end{eqnarray}
Neglecting the terms which are suppressed by $v_\Delta$ one has the simple forms \cite{Arhrib:2011uy,Akeroyd:2011ir}:
\begin{equation}
g_{H_1 H^{++}H^{--}}  \approx  - \lambda_1v \,,
\label{trilinHpp}
\end{equation}
\begin{equation}
g_{H_1 H^+H^-}  \approx  -(\lambda_{1} + \frac{\lambda_{4}}{2}) v \,.
\label{trilinHp}
\end{equation}
Therefore one expects that the couplings $g_{H_1 H^{++}H^{--}}$ and $g_{H_1 H^+H^-}$ have
similar magnitudes in the HTM, and it is essentially $\lambda_1$ which determines the value of
 $g_{H_1 H^{++}H^{--}}$. 

\begin{figure}[t]
\begin{center}
\includegraphics[origin=c, angle=0, scale=0.5]{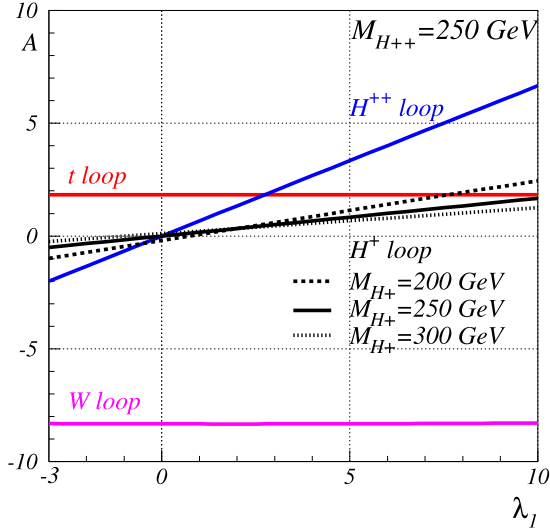}
\caption{The amplitudes ($A$) for the $W$, $t$, $H^{\pm\pm}$ and $H^{\pm}$
contributions to $H_1\to \gamma\gamma$ as a function of $\lambda_1$ for $m_{H^{\pm\pm}}=250$ GeV,
and $m_{H^\pm}=200$ GeV, 250 GeV and 300 GeV, with $m_{H_1}\sim 125$ GeV.}
\label{amp}
\end{center}
\end{figure}

In Fig.~\ref{amp} we plot the amplitudes (before squaring) of the separate contributions to 
$H_1\to \gamma\gamma$ in Eq.~(\ref{partial_width_htm}) for $m_{H_1}=125$ GeV, 
$m_{H^{\pm\pm}}$=250 GeV, and three values of $m_{H^\pm}$ (300 GeV, 250 GeV and 200 GeV).
Due to the enhancement factor of a factor of four from the
electric charge of $H^{\pm\pm}$ being twice that of $H^\pm$, one expects 
the contribution of $H^{\pm\pm}$ to $H_1\to \gamma\gamma$ to dominate that 
from $H^\pm$, and this can be seen in  Fig.~\ref{amp}.
In contrast, in the little Higgs model of
\cite{Han:2003gf} the theoretical structure of the model requires
 $g_{H_1 H^{++}H^{--}} << g_{H_1 H^+H^-}$, and so the contribution from $H^{\pm\pm}$ could be neglected
in \cite{Han:2003gf} with respect to the that from $H^\pm$. Note that the contribution from
the $H^{\pm\pm}$ loop interferes constructively with that of the $W$ loop for $\lambda_1<0$, while for
$\lambda_1>0$ the interference is destructive and its magnitude can be as large as that 
of the $W$ contribution  for $\lambda_1 \sim 10$. The $H^\pm$ loop is usually subdominant.
For $\lambda_4=0$ (i.e. $m_{H^{\pm\pm}}\sim m_{H^{\pm}}$) one has ${\tilde g}_{H_1 H^{++}H^{--}}={\tilde g}_{H_1 H^+H^-}$ and thus the loop from $H^\pm$ 
enhances the total amplitude from the charged scalars, with the contribution of the $H^\pm$ loop being roughly one quarter that of the
$H^\pm$ loop. If $\lambda_4$ and $\lambda_1$ have opposite signs and
$|\lambda_4/2| > |\lambda_1|$ then the contribution from $H^\pm$ interferes destructively with that
from $H^{\pm\pm}$.

\subsection{Theoretical constraints on $\lambda_1$ and $\lambda_4$}

Theoretical constraints on the parameters of the scalar potential of the HTM
come from the requirements of perturbative unitarity in scalar-scalar scattering, stability of the
potential, and positivity of the masses of the scalars.
A comprehensive study of all these constraints has been performed in \cite{Arhrib:2011uy},
with earlier studies in \cite{Dey:2008jm,Akeroyd:2010je}.
The magnitude of $\lambda_1$ plays a crucial role in determining the numerical value of 
the trilinear couplings in eq.(\ref{trilinHpp}) and eq.(\ref{trilinHp}). 
The main constraint on $\lambda_1$ comes from the requirement of the stability of the scalar
potential, and one of four such constraints derived in \cite{Arhrib:2011uy} is:
\begin{equation}
\lambda_1 + \sqrt{\lambda(\lambda_2+\frac{\lambda_3}{2})} > 0 \, .
\label{stability}
\end{equation}
If $\lambda_2$ and $\lambda_3$ are taken to be zero, then the 
combined constraints on $\lambda_1$ from perturbative unitarity in scalar-scalar scattering and from
stability of the potential require $\lambda_1 > 0$, as shown in \cite{Arhrib:2011uy}. However, for the case
of sufficiently positive $\lambda_2$ and $\lambda_3$ the choice of $\lambda_1 < 0$ is
permissible, and is a possibility which was acknowledged in \cite{Arhrib:2011uy} but its phenomenology was not studied.
We note that varying $\lambda_2$ and $\lambda_3$ has very little effect on the trilinear couplings
in eq.~(\ref{eq:ghHpp}) and eq.~(\ref{eq:ghHp}), and on the masses of the triplet scalars. In our numerical analysis we will
fix $\lambda_2=\lambda_3$, and use eq.(\ref{stability}) to determine $\lambda_2$ as a function of
$\lambda_1$ and $\lambda$.  For the case of $\lambda_4=0$ and $\lambda=0.516$ (on which we
will focus) one can show from eq~.(\ref{stability}) that the stability of the vacuum requires $\sqrt{0.77 \lambda_2}> -\lambda_1$.
The most negative value of $\lambda_1$ that we will consider is $\lambda_1=-3$, for which 
$\lambda_2=\lambda_3\sim 10$ would be necessary. For $\lambda_1=-2$ and $\lambda_1=-1$ one would
require $\lambda_2 >4$ and  $\lambda_2> 1$ respectively. 

\subsection{Limits on $m_{H^{\pm\pm}}$ from direct searches for $H^{\pm\pm}$} 

The strongest limits on $m_{H^{\pm\pm}}$ come from the ongoing LHC searches, which
assume production via  $\qqHppHmm$%
~\cite{Barger:1982cy, Gunion:1989in, Muhlleitner:2003me, Han:2007bk, Huitu:1996su}
and $\qqHpmpmHmp$~\cite{Barger:1982cy, Dion:1998pw, Akeroyd:2005gt, Perez:2008ha,delAguila:2008cj,Akeroyd:2010ip}.
Production mechanisms which depend on $v_\Delta$ 
(i.e.\ $pp\to W^{\pm *}\to W^\mp H^{\pm\pm}$
and fusion via 
$W^{\pm *} W^{\pm *} \to H^{\pm\pm}$%
~\cite{Huitu:1996su,Gunion:1989ci,Vega:1989tt})
have smaller cross sections than the above processes for
 $m_{H^{\pm\pm}}< 500\,\GeV$,
but such mechanisms could be the dominant source of 
$H^{\pm\pm}$ at the LHC
if $v_{\Delta}={\cal O}(1)\,\GeV$ and $m_{H^{\pm\pm}}> 500\,\GeV$.

 The first searches
for $H^{\pm\pm}$ at the LHC with $\sqrt s=7\,\TeV$
have been carried out by the CMS collaboration \cite{CMS-search}
(with $4.63\,\fb^{-1}$ of integrated luminosity).
Separate searches were performed
for $\qqHppHmm$ and $\qqHpmpmHmp$,
assuming the decay channels $H^{\pm\pm}\to \ell^\pm{\ell^\prime}^\pm$
and $\Hpmlpmnu$ with  $\ell=e,\mu,\tau$.
 The ATLAS collaboration has carried out three distinct searches
for the decay $H^{\pm\pm}\to \ell^\pm{\ell^\prime}^\pm$
(assuming production via $\qqHppHmm$ only) as follows:
 i) two (or more) leptons (for $\ell=\mu$ only),
using $1.6\,\fb^{-1}$ of integrated luminosity%
~\cite{Aad:2012cg};
 ii) three (or more) leptons ($\ell=e,\mu$),
using $1.02\,\fb^{-1}$ of integrated luminosity%
~\cite{ATLAS-search:3l};
 iii) four leptons ($\ell=e,\mu$),
using $1.02\,\fb^{-1}$ of integrated luminosity~\cite{ATLAS-search:4l}.
The mass limits on $m_{H^{\pm\pm}}$ from the LHC searches are
stronger than those from the Tevatron searches \cite{Aaltonen:2011rta,Abazov:2011xx}.
In the CMS search (which currently has the strongest limits),
the limit is $m_{H^{\pm\pm}}> 450\,\GeV$ for the decay with $\ell=e,\mu$ and assuming BR=100\% in a given channel. For final states with one $\tau$ or two $\tau$ 
the limits are weaker, being $m_{H^{\pm\pm}}> 350\,\GeV$ and $m_{H^{\pm\pm}}> 200\,\GeV$
respectively. Moreover, the limit of $m_{H^{\pm\pm}}>400$ GeV is derived in four benchmark points in the HTM in which all
six decay channels have a non-zero BR($H^{\pm\pm}\to \ell^\pm{\ell^\prime}^\pm$).

The above searches all assume dominance of the leptonic decay channels $H^{\pm\pm}\to \ell^\pm{\ell^\prime}^\pm$,
which is the case in the HTM for  $v_\Delta < 10^{-4}\,\GeV$ and degeneracy of the triplet scalars.
However, the decay channel $H^{\pm\pm}\to W^\pm W^\pm$ dominates for $v_\Delta > 10^{-4}\,\GeV$
(and assuming degeneracy of the triplet scalars), for which
there have been no direct searches. Therefore the above bounds on $m_{H^{\pm\pm}}$ cannot be applied to the 
case of  $v_\Delta > 10^{-4}\,\GeV$, and in this scenario $H^{\pm\pm}$ could be much lighter. 
Moreover, for the case of non-degeneracy of the triplet scalars the decay $H^{\pm\pm}\to H^\pm W^*$
can be dominant over a wide range of values of $100 \,{\rm eV} <v_\Delta < 1$ GeV
\cite{Chakrabarti:1998qy,Chun:2003ej,Akeroyd:2005gt,Perez:2008ha,Akeroyd:2011zz,Melfo:2011nx,Aoki:2011pz}, and thus the limits on 
$m_{H^{\pm\pm}}$ from the searches for $H^{\pm\pm}\to \ell^\pm{\ell^\prime}^\pm$ can be weakened.
In our numerical analysis we will consider values of $m_{H^{\pm\pm}}$ as low as 150 GeV.

\subsection{Searches for $H \to \gamma\gamma$ at the LHC} 
Two LHC collaborations have performed inclusive searches for $H\to \gamma\gamma$ 
(CMS in \cite{Chatrchyan:2012tw} and ATLAS in \cite{ATLAS:2012ad}).
Assuming that the production cross section is that of the SM Higgs boson, 
constraints are now being derived on the quantity BR$(H \rightarrow \gamma\gamma)$
in any theoretical model:
\begin{equation}
R_{\gamma\gamma}=\frac{{\rm BR}(H \rightarrow \gamma\gamma)^{MODEL}} 
{{\rm BR}(H \rightarrow \gamma\gamma)^{SM}}\,.
\label{eq:RSM}
\end{equation}
The dominant production process in the SM
is $gg\to H$ (which comprises around 87\% of the inclusive production of $H$)
while vector boson fusion $qq\to qq W^*W^*\to Hjj$, Higgsstrahlung
$qq\to W^*/Z^* \to HW/HZ$ and $gg\to Ht\overline t$ give subdominant contributions.
 Moreover, both CMS \cite{CMS-ferm} and ATLAS \cite{Aad:2012yq} 
have performed searches for a fermiophobic Higgs boson decaying to two photons.
These latter searches apply selection cuts which differ from those in \cite{Chatrchyan:2012tw,ATLAS:2012ad}.
This is to increase sensitivity to the vector boson fusion and the Higgsstrahlung production mechanisms
and to suppress any contribution from $gg\to H$, which is expected to be very small or even absent 
for a fermiophobic Higgs boson.
In \cite{Giardino:2012ww} it is estimated that the search strategy in \cite{CMS-ferm}
is sensitive to the combination
\begin{equation}
0.033\sigma(gg\to H)+ \sigma(qq\to Hjj)\times {\rm BR}(H\to \gamma\gamma)\,,
\end{equation}
while the search strategy in \cite{Aad:2012yq}  is sensitive to the combination
\begin{equation}
0.3\sigma(gg\to H)+ \sigma(qq\to WH, ZH, Hjj)\times {\rm BR}(H\to \gamma\gamma)\,.
\end{equation}
In all four of these searches for $H\to \gamma\gamma$ there is a small excess of events at $\sim 125$ GeV, which leads
to a weaker exclusion of $R_{\gamma\gamma} > 3.5$ at 95\% c.l.
The small excess could merely be a background fluctuation which will disappear with a larger integrated luminosity.
However, attributing this excess to genuine production of $H\to \gamma\gamma$ \cite{Giardino:2012ww} gives rise to the
following best fit value of $R_{\gamma\gamma}$, which is derived from averaging the four measurements in 
\cite{Chatrchyan:2012tw, ATLAS:2012ad,CMS-ferm,Aad:2012yq}:
\begin{equation}
R_{\gamma\gamma}=2.1\pm 0.5\,.
\end{equation}
Away from $m_H=125$ GeV, values of $R_{\gamma\gamma}$ in the interval $1 < R_{\gamma\gamma} < 2$ are excluded for
$110\, {\rm GeV} < m_H  < 150$ GeV, with the sensitivity reaching $R_{\gamma\gamma}\sim 1$ in some
very small mass intervals.

In the HTM the couplings of $H_1$ to the quarks and vector bosons are almost identical to those
of the SM Higgs boson, and are given in Eqs~.(\ref{h1tt}) to (\ref{h1ZZ}).
Since $\cos\alpha \sim 1$ and  $\cos\beta'\sim 1$  then the production rate
of $H_1$ at the LHC is essentially the same as that of the SM Higgs boson in all the standard search channels.
As in \cite{Arhrib:2011uy} we define the simple ratio: 
\begin{equation}
R_{\gamma\gamma}=\frac{(\Gamma(H_1\rightarrow gg)
\times {\rm BR}(H_1\rightarrow \gamma\gamma))^{HTM}}
{(\Gamma(H\rightarrow gg)\times 
{\rm BR}(H \rightarrow \gamma\gamma))^{SM}} \, .
\label{eq:Rgg}
\end{equation}
This is approximately the quantity which is now being constrained by the LHC searches
in \cite{Chatrchyan:2012tw} and in \cite{ATLAS:2012ad}, for which the dominant
production mechanism comes from $gg\to H_1\to \gamma\gamma$.
Given the coupling in eqs.~(\ref{h1tt}) and (\ref{h1bb}), one can see that the ratio 
$\Gamma(H_1\rightarrow gg)/\Gamma(H \rightarrow gg)$ is essentially equal to one in the HTM.
As mentioned above, the fermiophobic Higgs searches in \cite{CMS-ferm} and ATLAS \cite{Aad:2012yq} 
greatly reduce the contribution from $gg\to H_1$. For these searches one would need to replace $\Gamma(H_1\rightarrow gg)$ in 
eq.~(\ref{eq:Rgg}) by the couplings to vector bosons in eq.~(\ref{h1WW}), and the production rate for $H_1$ in the HTM
would still be essentially the same as that of the SM Higgs boson.

\section{Numerical Analysis} 
In this section we quantify the magnitude of the charged scalar loops
($H^{\pm\pm}$ and $H^\pm$) on the ratio 
$R_{\gamma\gamma}$. The case of $\lambda_1>0$ was studied in
\cite{Arhrib:2011uy}. We confirm their results, and present results
for the case of $\lambda_1<0$.
In our numerical analysis the following parameters are fixed:
\begin{eqnarray}
\lambda=0.516, \,\,\,{\rm leading \,to} \,\,\, m_{H_1}\sim 125 \,{\rm GeV} \, ;\\
v_\Delta=0.1 {\rm GeV} \, ; \\
\lambda_2=\lambda_3 \, .
\label{input-param}
\end{eqnarray}
The choice of $m_{H_1}\sim 125$ GeV is motivated by the excess of events
at this mass in the searches for $H_1\to \gamma\gamma$ at the LHC.
The value of $v_\Delta=0.1$ GeV satisfies the constraints on  $v_\Delta$ 
from the $\rho$ parameter.
Moreover, this choice of $v_\Delta$ ensures that BR($H^{\pm\pm}\to \ell^\pm\ell^\pm$) is negligible, and so
the strong limit of $m_{H^{\pm\pm}}>400$ GeV for the benchmark points in the HTM does not apply due to the dominance of
the decay mode $H^{\pm\pm}\to WW$ (or $H^{\pm\pm}\to H^\pm W^*$ if there is a mass splitting
between $H^{\pm\pm}$ and $H^\pm$). We emphasise that the role of $\lambda_2$ and $\lambda_3$
in our numerical analysis is merely to ensure that 
the constraint from vacuum stability in eq.~(\ref{stability}) is satisfied
for the novel case ($\lambda_1 <0$) of interest to us.
Our numerical results are essentially insensitive to the choice of $\lambda_2(=\lambda_3)$,
and we have explicitly checked that taking $\lambda_2=10$ (which is necessary to satisfy eq.~(\ref{stability})
with $\lambda_1=-3$) gives practically the same numerical results as 
for $\lambda_2=0$ for our choice of $v_\Delta=0.1$ GeV.
This is because $\lambda_2$ and $\lambda_3$ have an almost negligible effect on the trilinear couplings $H_1 H^{++}H^{--}$ 
and $H_1 H^+H^-$ in eq.~(\ref{eq:ghHpp}) and eq.~(\ref{eq:ghHp}), a consequence of the fact that
their contribution is multiplied by the small parameter $v_\Delta$. For definiteness we take $\lambda_2=\lambda_3=0.2$.
The parameter $\lambda_4$ determines whether there 
is a mass splitting among the triplet scalars, and $\lambda_4$ also enters 
the expression for the $H_1 H^+H^-$ coupling in Eq.~(\ref{eq:ghHp}).
We treat $\lambda_1$ and $m_{H^{\pm\pm}}$ as free parameters that
essentially determine the magnitude of
the $H^{\pm\pm}$ contribution to $H_1\to \gamma\gamma$.
We present results for the range:
\begin{eqnarray}
-3 < \lambda_1 < 10\,; \;\;\;  150\, {\rm GeV} < m_{H^{\pm\pm}} < 600\, {\rm GeV} \, .
\label{range}
\end{eqnarray}
We note that the range $0 < \lambda_1 < 10$ was studied in \cite{Arhrib:2011uy}.

\begin{figure}[t]
\begin{center}
\includegraphics[origin=c, angle=0, scale=0.5]{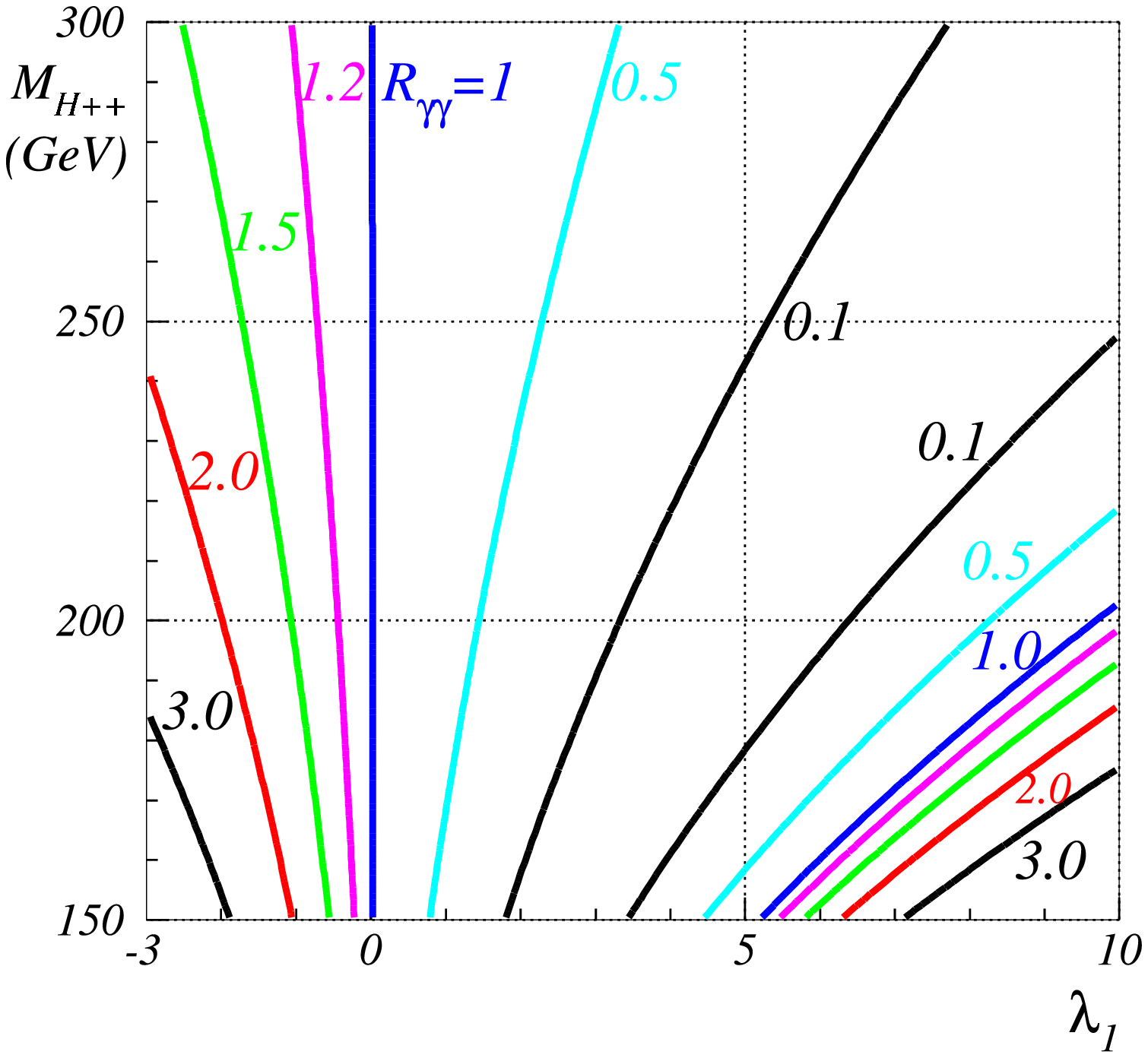}
\includegraphics[origin=c, angle=0, scale=0.5]{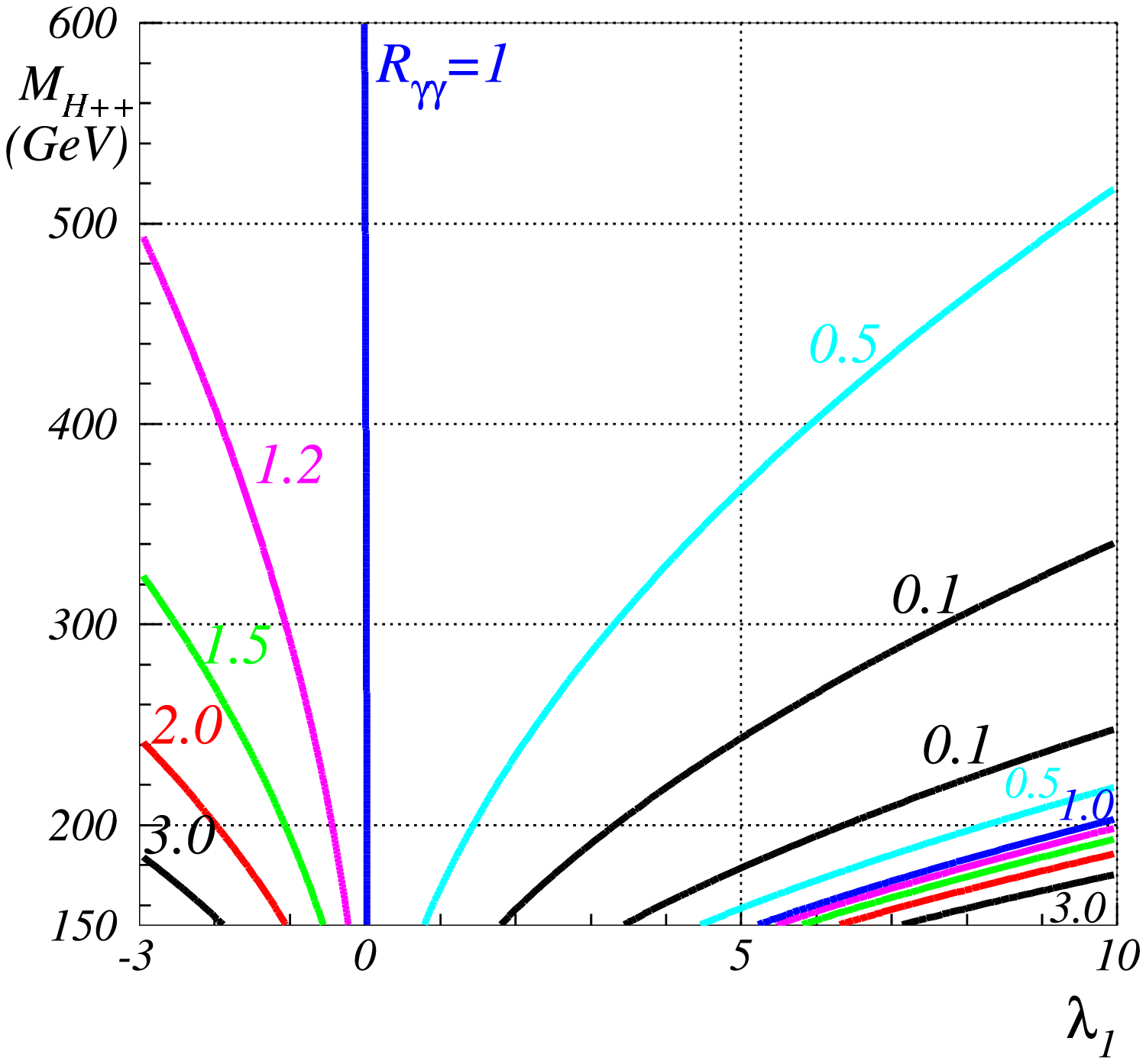}
\caption{The ratio $R_{\gamma\gamma}$ in the plane of $[\lambda_1, m_{H^{\pm\pm}}]$
for $150\,{\rm GeV} < m_{H^{\pm\pm}} < 300$ GeV (left panel) and
$150\,{\rm GeV} < m_{H^{\pm\pm}} < 600$ GeV (right panel), for $m_{H_1}\sim 125$ GeV and
$m_{H^{\pm\pm}}=m_{H^\pm}$.}
\label{rgam1}
\end{center}
\end{figure}

In Fig.~\ref{rgam1}, $R_{\gamma\gamma}$ is plotted in the plane of
$[\lambda_1, m_{H^{\pm\pm}}]$. We fix  $\lambda_4=0$ and so $H^{\pm\pm}$ and $H^{\pm}$ are 
essentially degenerate, and the couplings $g_{H_1H^{++}H^{--}}$ and $g_{H_1H^+H^-}$
in eq.~(\ref{trilinHpp}) and eq.~(\ref{trilinHp}) are approximately equal. Therefore the
contributions of $H^{\pm\pm}$ and $H^{\pm}$ to the decay rate of $H_1\to \gamma\gamma$
in eq.~(\ref{partial_width_htm}) differ only by their electric charge, with the
contribution of $H^{\pm\pm}$ being four times larger at the amplitude level.
In the left panel the range 
$150\, {\rm GeV} < m_{H^{\pm\pm}} < 300$ GeV is plotted, and in the right panel
the range $150\, {\rm GeV} < m_{H^{\pm\pm}} < 600$ GeV is plotted.
For the case of $\lambda_1>0$ one has destructive interference of the $H^{\pm\pm}$ loop
with that of the $W$ loop, leading to a significant suppression of $R_{\gamma\gamma}$ i.e.
in the region $0< \lambda_1 < 5$ and $150 \,{\rm GeV}< m_{H^{\pm\pm}}<300$ GeV there is a large parameter space 
for $R_{\gamma\gamma}<0.5$. For $0< \lambda_1 < 5$ and $400 \,{\rm GeV}< m_{H^{\pm\pm}}<600$ GeV
(i.e. the mass region which has yet to be probed in the direct searches which 
assume dominance of the decay $H^{\pm\pm}\to \ell^\pm\ell^\pm$)
the suppression is more mild, with $0.5 < R_{\gamma\gamma} < 1$. 
Consequently, for $0< \lambda_1 < 5$ a statistically significant signal for $H_1\to \gamma\gamma$ would
require considerably more integrated luminosity than for the case of the Higgs boson of the SM, and so
detection of $H_1\to \gamma\gamma$ might not be possible in the 8 TeV run of the LHC. However,
the other LHC search channels which make use of the tree-level decays (i.e $H_1 \to WW, ZZ$ etc)
would have the same detection prospects as those of the Higgs boson of the SM. We note that the
scenario of a large mixing angle $\sin\alpha \sim \pi/4$ \cite{Dey:2008jm,Akeroyd:2010je,Arhrib:2011uy} (for which $m_{H_1}\sim m_{H_2}$ is necessary) 
could further delay detection of
$H_1$ in these latter channels because the production cross sections of $H_1$ (and $H_2$) would have a suppression 
factor of $1/2$ relative to that of the Higgs boson of the SM, as well as having masses which differ by
a few GeV.

The case of  $R_{\gamma\gamma}=1$ occurs for $\lambda_1\sim 0$
(i.e. a negligible trilinear coupling $H_1 H^{++}H^{--}$), and also for 
a straight line which joins the points $\lambda_1=5$, $m_{H^{\pm\pm}}=150$ GeV
and $\lambda_1=10$, $m_{H^{\pm\pm}}=200$ GeV. Hence any signal
for $H_1\to \gamma\gamma$ with  $R_{\gamma\gamma}\sim 1$ 
(and assuming $m_{H^{\pm\pm}}<600$ GeV) would
restrict the parameter space of $[\lambda_1, m_{H^{\pm\pm}}]$ to two regions: i)
the region of $\lambda_1\sim 0$, and ii) the region of $\lambda_1 > 5$.
If $H^{\pm\pm}$ is very heavy and out of the discovery reach of the LHC (e.g. $m_{H^{\pm\pm}}>>1$ TeV) then
$R_{\gamma\gamma}\sim 1$ could be accommodated with any positive and sizeable $\lambda_1$.
As emphasised in \cite{Arhrib:2011uy}, 
in the region of $5< \lambda_1 < 10$ and  $m_{H^{\pm\pm}}<200$ GeV the contribution of the $H^{\pm\pm}$
loop is so large that $R_{\gamma\gamma}>1$ occurs. This region is still compatible
with current LHC data, which excludes $R_{\gamma\gamma}>3.5$ for $m_{H_1}$ around 125 GeV, 
while $R_{\gamma\gamma}> 2$ is excluded for essentially all other choices of
$m_{H_1}$ in the interval $110 \,{\rm GeV} < m_{H_1}< 150$ GeV.
The excess of $\gamma\gamma$ events at 125 GeV, if assumed to originate from a Higgs boson,
roughly corresponds to $R_{\gamma\gamma}=2.1\pm 0.5$. If $R_{\gamma\gamma}>1$ 
turns out to be preferred by LHC data then one interpretation in the HTM would be
the region of  $5< \lambda_1 < 10$ and  $m_{H^{\pm\pm}}<200$ GeV \cite{Arhrib:2011uy}.

We now discuss the case of $\lambda_1<0$, for which $R_{\gamma\gamma}>1$.
The current sensitivity to $H_1\to \gamma\gamma$ in the LHC searches is
between $1< R_{\gamma\gamma}< 2$ in the mass range $110 < m_{H_1} < 150$ GeV, and 
so the scenario of $\lambda_1<0$ is now being probed by the ongoing searches.
One can see that the current best fit value of $R_{\gamma\gamma}=2.1\pm 0.5$ 
can be accommodated by values of $|\lambda_1|$ which are much smaller than for the case of
$\lambda_1 >0$, e.g. $R_{\gamma\gamma}=2$ can be obtained for $\lambda_1\sim -1$ (or $\lambda_1\sim 6$) 
and $m_{H^{\pm\pm}}=150$ GeV. Importantly, any measured value of $R_{\gamma\gamma}>1$ would
be readily accommodated by the scenario $\lambda_1 < 0$, even for a relatively heavy $H^{\pm\pm}$,
e.g. for $m_{H^{\pm\pm}}>400$ GeV one has
$1.0 < R_{\gamma\gamma}< 1.3$. If either of the
decays $H^{\pm\pm}\to WW$ or $H^{\pm\pm}\to H^\pm W^*$ is dominant (for which there have been
no direct searches) then $m_{H^{\pm\pm}}<400$ GeV is not experimentally excluded, and larger
enhancements of  $R_{\gamma\gamma}$ are possible 
e.g.  $R_{\gamma\gamma}= 4.5, 3.1$ and 1.9 for  $\lambda_1=-3, -2$ and $-1$, with $m_{H^{\pm\pm}}=150$ GeV.
Such large enhancements of $R_{\gamma\gamma}$ would require a relatively light $H^{\pm\pm}$
(e.g. $m_{H^{\pm\pm}}<300$ GeV) which decays dominantly to $H^{\pm\pm}\to WW$ and/or $H^{\pm\pm}\to H^\pm W^*$,
for which there have been no direct searches. Simulations of $H^{\pm\pm}\to WW$ were performed in
\cite{Perez:2008ha,Chiang:2012dk}, with good detection prospects for $m_{H^{\pm\pm}}<300$ GeV.
A parton-level study of $H^{\pm\pm}\to H^\pm W^*$ (for the signal only) has been carried out in
\cite{Aoki:2011pz}.

\begin{figure}[t]
\begin{center}
\includegraphics[origin=c, angle=0, scale=0.5]{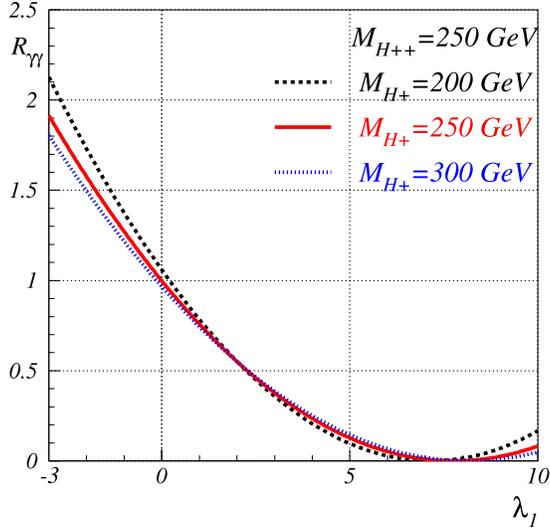}
\caption{The ratio $R_{\gamma\gamma}$ as a function of $\lambda_1$ for $m_{H^{\pm\pm}}=250$ GeV,
and $m_{H^\pm}=200$ GeV, 250 GeV and 300 GeV, with $m_{H_1}\sim 125$ GeV.}
\label{rgam-hp}
\end{center}
\end{figure}

The above discussion was for the case of $\lambda_4=0$ and we now discuss the effect
of $\lambda_4\ne 0$. In this case the couplings $g_{H_1H^{++}H^{--}}$ and $g_{H_1H^+H^-}$
in eq.~(\ref{trilinHpp}) and eq.~(\ref{trilinHp})
are not equal (as discussed in Fig.~\ref{amp}), and there is a mass splitting between  $m_{H^{\pm\pm}}$ and
$m_{H^{\pm}}$ given by eq.~(\ref{charged-mass}) and eq.~(\ref{doub-charged-mass}).
Therefore the contribution of $H^{\pm\pm}$ to $H_1\to \gamma\gamma$ is not simply 
four times the contribution of $H^\pm$ at the amplitude level. 
In Fig.~\ref{rgam-hp} we plot $R_{\gamma\gamma}$ as a function of $\lambda_1$ , fixing
$m_{H^{\pm\pm}}=250$ GeV and taking $m_{H^\pm}=200$ GeV, 250 GeV and 300 GeV (corresponding to
$\lambda_4=-1.48, 0$ and $1.82$ respectively). One can see that the case 
of $m_{H^\pm}=200$ GeV and $\lambda_1 < 0$ leads to a value of $R_{\gamma\gamma}$
which is roughly $10\%$ larger than the value for the case of $m_{H^\pm}=m_{H^{\pm\pm}}$. This
due to the an increase of the magnitude of the coupling $g_{H_1 H^+H^-}$ in eq.~(\ref{trilinHp}) and
also because of less suppression from the $1/m^2_{H^\pm}$ term in eq.~(\ref{eq:redgcalHHp}). Note that
$R_{\gamma\gamma}\ne 1$ when $\lambda_1=0$ due to the non-vanishing coupling $g_{H_1 H^+H^-}$ for
$\lambda_4\ne 0$ (see also Fig.~(\ref{amp})). For $\lambda_1 > 0$ the magnitude of $g_{H_1 H^+H^-}$ is 
less than that of $g_{H_1 H^{++}H^{--}}$ due to destructive interference in  eq.~(\ref{trilinHp}), and at around
$\lambda_1=2$ the value of $R_{\gamma\gamma}$ becomes equal to the case of $m_{H^\pm}=m_{H^{\pm\pm}}$. For
$\lambda_1 > 8$ one finds again values of $R_{\gamma\gamma}$ which are slightly larger than for the case
of  $m_{H^\pm}=m_{H^{\pm\pm}}$. For $m_{H^\pm}=300$ GeV (i.e. $\lambda_4=1.82$) the converse dependence 
of $R_{\gamma\gamma}$ on $\lambda_1$ is found.

\section{Conclusions}
Ongoing searches at the LHC for the decay $H\to \gamma\gamma$ are approaching 
sensitivity to the prediction for the SM Higgs boson in the range $110 \,{\rm GeV} < m_H < 150$ GeV. This data
constrains models of New Physics which enhance BR($H\to \gamma\gamma$) relative to the rate in the SM. 
A local excess of events is seen by both the CMS and ATLAS experiments, which is consistent with a signal for the
Higgs boson of the SM with a mass of around 125 GeV.
Doubly charged Higgs bosons ($H^{\pm\pm}$), which arise in the Higgs Triplet Model (HTM) of neutrino mass generation,
can significantly alter the branching ratio of $H_1\to \gamma\gamma$ (where $H_1$ is the lightest CP-even scalar in the HTM), 
while the branching ratios of the dominant tree-level decays are essentially the same as those for the Higgs boson of the SM.
The contribution of the loops involving $H^{\pm\pm}$ is mediated by a trilinear coupling
$H_1 H^{++} H^{--}$, and its magnitude 
essentially depends on two parameters: an arbitrary quartic coupling
($\lambda_1$) and the mass of $H^{\pm\pm}$ ($m_{H^{\pm\pm}}$). Consequently, the 
improving limits on BR($H_1\to \gamma\gamma$) from LHC data constrain the parameter space of $[\lambda_1, m_{H^{\pm\pm}}]$.
An additional (and subdominant) contribution comes from a loop with $H^\pm$, and its magnitude
is determined by $\lambda_1$, $\lambda_4$ and $m_{H^\pm}$.
As recently pointed out in \cite{Arhrib:2011uy}, for the case of $\lambda_1>0$ the
contribution of $H^{\pm\pm}$ interferes destructively with that of the $W$ loop, and it can lead to
a significant suppression of the branching ratio of 
BR($H_1\to \gamma\gamma$). In this scenario much more data would be
required to see a signal for $H_1\to \gamma\gamma$ in the HTM at the LHC.
For very large positive $\lambda_1$ (i.e. $\lambda_1>5$) and $m_{H^{\pm\pm}}<200$ GeV
a sizeable enhancement of BR($H_1\to \gamma\gamma$) would be possible, and this parameter space
is now being constrained by the LHC limits on BR($H_1\to \gamma\gamma$), as well as being a possible
explanation of the excess of events at 125 GeV  \cite{Arhrib:2011uy}.

In this work we pointed out that constructive interference of the $H^{\pm\pm}$ contribution with the
$W$ contribution occurs for $\lambda_1<0$, and such a parameter space is consistent with
theoretical constraints on $\lambda_1$ from requiring the stability of the vacuum of the scalar potential. For $m_{H^{\pm\pm}}=400$ GeV, which
is roughly the bound if the decays $H^{\pm\pm}\to \ell^\pm\ell^\pm$ decays are dominant in the HTM, an enhancement
of up to $\sim 1.3,1.2$ and $1.1$ is possible for $\lambda_1=-3,-2$ and $-1$. Conversely, if either of the
decays $H^{\pm\pm}\to WW$ or $H^{\pm\pm}\to H^\pm W^*$ is dominant (for which there have been
no direct searches) then $m_{H^{\pm\pm}}<400$ GeV is not experimentally excluded, and larger
enhancements of $\sim 4.5, 3.1$ and $1.9$ are possible for  $\lambda_1=-3,-2$ and $-1$ with $m_{H^{\pm\pm}}=150$ GeV.
Consequently, the parameter space of $\lambda_1 <0$ in the HTM 
is more tightly constrained by the ongoing searches for $H_1 \to \gamma\gamma$ than the case of $\lambda_1 > 0$. 
Importantly, the case of $\lambda_1 <0$ would readily accommodate any signal for $H_1\to \gamma\gamma$ with 
a branching ratio which is higher than that for the Higgs boson, 
for {\it smaller} values of $|\lambda_1|$ than for the case of $\lambda_1>0$. 
In such a scenario, dedicated searches at the LHC for the decay channels $H^{\pm\pm}\to WW$ or $H^{\pm\pm}\to H^\pm W^*$ with
$m_{H^{\pm\pm}}<400$ GeV would be strongly motivated.

\section*{Acknowledgements}
 A.G.A was supported by a Marie Curie Incoming International Fellowship,
 FP7-PEOPLE-2009-IIF, Contract No. 252263.
SM is partially supported through the NExT Institute.

\end{document}